\documentstyle[twocolumn,floats,epsf,prl,aps]{revtex}

\begin{document}

\draft
\title{Oscillatory Behavior of the Rate of Escape through an Unstable Limit
Cycle}
\author{Robert S. Maier${}^{(1)}$ and D.~L. Stein${}^{(2)}$}
\address{Mathematics${}^{(1)}$ and Physics${}^{(2)}$ Departments, University of
Arizona, Tucson, Arizona 85721}

\maketitle

\begin{abstract}
Suppose a two-dimensional dynamical system has a stable attractor that is
surrounded by an unstable limit cycle.  If~the system is additively
perturbed by white noise, the rate of escape through the limit cycle will
fall~off exponentially as the noise strength tends to zero.  By~analysing
the associated Fokker--Planck equation we~show that in~general, the
weak-noise escape rate is non-Arrhenius: it~includes a factor that is
periodic in the logarithm of the noise strength.  The presence of this
slowly oscillating factor is due to the nonequilibrium potential of the
system being nondifferentiable at the limit cycle.  We point~out the
implications for the weak-noise limit of stochastic resonance models.
\end{abstract}

\pacs{PACS numbers: 02.50.-r, 05.40.+j}

\narrowtext

A particularly interesting phenomenon is the occurrence of noise-induced
transitions between attractors of a dynamical system.  Such transitions
occur in chemical physics, where the transition is a motion across a
transition-state surface from a reactant region to a product region.  They
also occur in statistical physics, and in other fields where stochastic
modelling plays a role~\cite{basics,Naeh90}.

If~the noise is white, or has a short correlation time and may be
approximated as white, then the probability density of the system will
satisfy a~Fokker--Planck equation.  This equation governs the way in~which
noise-induced transitions occur.  By~the `rate' at~which a specified
transition takes place we~shall mean the probability that it~occurs, per
unit time.  At~least in finite-dimensional systems, any such rate should
fall~off exponentially as~$\epsilon$, the noise strength, tends to zero.
(In~thermal applications $\epsilon$~would be proportional to~$kT$.)
In~fact, each transition should be characterized by an activation
energy~$\Delta W$, with the transition rate falling~off to leading order as
$e^{-\Delta W/\epsilon}$.  Computing the pre-exponential factor requires a
careful analysis of the Fokker--Planck
equation~\cite{Naeh90,MaierB,MaierCF,Talkner87}.

Most work has focused on the case when the competing attractors of the
dynamical system are separated by a separatrix ({\em i.e.},~a `ridge')
containing a saddle point.  However, models where the separatrix is instead
an unstable limit cycle arise in the context of chemical reactions
constrained to occur far from equilibrium \cite{chemphys}.  Also,
transitions across an unstable limit cycle separating steady states of
periodic vibration occur in models of stochastic resonance in bistable
continuous systems \cite{Dykman79,Dykman95}.  A~full analysis of
noise-driven escape through an unstable limit cycle accordingly seems
called~for.

Previous work on noise-driven transitions in models with an unstable limit
cycle is found in Refs.~\cite{Naeh90,Graham84,Graham85,Day90,Day93,Day93a}.
Graham and T\'el~\cite{Graham84,Graham85}, and Day
\cite{Day90,Day93,Day93a}, have noted that the nonequilibrium
potential~$W$, as a function on the state space of the system being
modelled, will be nondifferentiable (`wild') near the limit cycle if the
system fails to satisfy a form of detailed balance.  Naeh
{\em et~al.}~\cite{Naeh90} began an analysis of the Fokker--Planck equations
associated to such models, by a method of matched asymptotic
expansions, but their analysis assumed that $W$
was differentiable at the limit cycle.

In this Letter we begin an asymptotic analysis of the rate of escape
through an unstable limit cycle that incorporates the insights of Graham
and T\'el, and of Day, and obtain a striking result.  We~show that
generically, in two-dimensional models with an unstable limit cycle
enclosing an attractor, the rate of escape~$R$ is given by a
non-Arrhenius formula of the form
\begin{equation}
\label{eq:slow}
R\sim \mbox{const}
\times\epsilon^b G\left(\left|\log\epsilon\right|\right) e^{-\Delta
W/\epsilon}
\end{equation}
in the weak-noise ($\epsilon\to0$) limit.  Here $b$~is model-dependent, and
the factor $G\left(\left|\log\epsilon\right|\right)$ is a model-dependent
periodic function of~$\left|\log\epsilon\right|$.  The presence of such
{\em slowly oscillating factors\/} in the expressions for noise-dependent
transition rates has not previously been suspected.  It~indicates that even
in bistable dynamical systems with effective dimensionality as low as two,
relaxation phenomena may be more complicated than is commonly believed.
Our analysis applies whenever the system is truly two-dimensional, {\em
i.e.}, is nonseparable.

{\em Models\/}.---We consider models with dynamics that are those of a
Brownian particle moving in a drift field, {\em i.e.},
\begin{equation}
\dot x^i = u^i(\bbox{x}) 
+ \epsilon^{1/2} \sum_{\alpha=1}^2 {\sigma^i}_\alpha(\bbox
{x})\eta_\alpha(t).
\end{equation}
Here $\bbox {x} = (x^1,x^2)$ is a pair of state variables, and the drift
field $\bbox {u}=(u^1,u^2)$ specifies the dynamics in the absence of noise.
$(\eta_1,\eta_2)$ is a pair of white noise processes, satisfying $\langle
\eta_\alpha(s)\eta_\beta(t)\rangle = \delta_{\alpha\beta}\delta(s-t)$.
$\bbox {\sigma}=({\sigma^i}_\alpha)$ is a so-called noise matrix that is
allowed to be state-dependent (a~`zweibein' field).  The probability
density $\rho=\rho(\bbox {x},t)$ of such a system satisfies the
Fokker--Planck equation
\begin{equation}
\dot \rho = -{\cal L}_\epsilon^*\rho \equiv (\epsilon/2)
\partial_i\partial_j\left[D^{ij}(\bbox {x})\rho\right]
- \partial_i\left [u^i(\bbox {x})\rho\right],
\end{equation}
where the diffusion tensor $\bbox {D}=(D^{ij})\equiv
{\bbox{\sigma}}{\bbox{\sigma}}^t$.  The operator ${\cal L}_\epsilon^*$ is
the (forward) Fokker--Planck operator.  We~consider here the case when
there is a point attractor~$S$ in the $(x^1,x^2)$-plane, with domain of
attraction~$\Omega$, for~which the boundary~$\partial\Omega$ is an unstable
limit cycle.

\begin{figure}
\centering
\epsfxsize=2.3in
\leavevmode\epsfbox{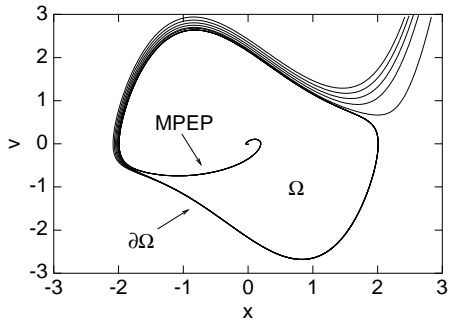}
\caption{The unstable limit cycle $\partial\Omega$ of the van der Pol
model, and the MPEP\null, which emerges from the attractor~$(0,0)$ and
spirals into~it.  The trajectories exiting from~$\Omega$ are optimal
trajectories that are perturbations of the MPEP\null.}
\label{fig:trajs}
\end{figure}

This framework is sufficiently general that it can accomodate
two-dimensional models with overdamped dynamics, or one-dimensional models
with underdamped dynamics.  In~the latter case one of the state variables
($x^1$,~say) would be a position, and the other a velocity.  Our
simulations below are of a model of this sort, namely
\begin{eqnarray}
\label{eq:vanderPol1}
\dot x &=& v\\
\dot v &=& -x + (v^2-1) v  + \epsilon^{1/2}\eta_2(t).
\label{eq:vanderPol2}
\end{eqnarray}
This is a time-reversed van~der Pol oscillator, the asymptotic analysis of
which was begun by Day~\cite{Day93a}.  Here $(x^1,x^2)=(x,v)$, $S=(0,0)$,
and $\bbox {D}={\mbox{diag}}\,(0,1)$ is degenerate.  The unstable limit
cycle is shown in Fig.~\ref{fig:trajs}.

{\em The Analysis\/}.---To estimate the rate of escape through~$\partial
\Omega$, we use the Kramers flux-over-the-barrier
technique~\cite{Kramers40}.  Suppose that escaping Brownian particles are
re-injected at the attractor~$S$, and a steady state has been reached.  The
probability density in this state, which we denote~$\rho_0$, will satisfy
${\cal L}_\epsilon^*\rho_0 = 0$.  When the noise strength~$\epsilon$ is
small, $\rho_0$~will be tightly peaked near~$S$.  $\rho_0(\bbox {x})$, at
points~$\bbox {x}$ on the limit cycle~$\partial\Omega$ and outside~it,
will be suppressed by a factor $\sim e^{-\Delta W/\epsilon}$ relative to
$\rho_0(S)$.  Since the Fokker--Planck equation has the form of a
continuity equation, with current density $J^i[\rho]$ equalling $\rho u^i -
(\epsilon/2)\partial_j\left[D^{ij}\rho\right]$, the escape rate~$R$ may be
computed as the flux of probability through~$\partial\Omega$, {\em i.e.},
\begin{equation}
\label{eq:flux}
R \sim \left.\int_{\partial \Omega} \bbox {J}[\rho_0]\cdot \bbox {n}\,d\ell
\right/\int_\Omega \rho_0\,d^2x.
\end{equation}
Here $\bbox {n}$ denotes the outward normal on~$\partial\Omega$.

To~derive the oscillating formula of eq.~(\ref{eq:slow}) from
eq.~(\ref{eq:flux}), we~introduce a WKB approximation to the steady-state
density~$\rho_0$ in the weak-noise
limit~\cite{Naeh90,MaierB,MaierCF,Talkner87}.  We~write
\begin{equation}
\label{eq:WKB}
\rho_0 (\bbox {x}) \sim K(\bbox {x}) \exp\left(- 
W(\bbox {x})/\epsilon\right).
\end{equation}
Here $W(\bbox {x})$ is an `activation energy' controlling noise-induced
fluctuations from the attractor to the vicinity of~$\bbox {x}$.  Though
(\ref{eq:WKB}) resembles a Maxwell-Boltzmann distribution, $W$~is a {\em
nonequilibrium\/} potential, since the steady state is not necessarily an
equilibrium state, in~that it~does not necessarily satisfy detailed
balance.  We~set $W(S)=0$, so~that $\Delta W$, the falloff rate of the
escape rate, is the value of~$W$ attained on the unstable limit cycle.

Substituting (\ref{eq:WKB}) into ${\cal L}^*_\epsilon \rho_0=0$ and
separating~out the ${\cal O}(\epsilon^{-1})$ terms yields the eikonal
equation
\begin{equation}
\label{eq:eikonal}
H(x^i, \partial W/\partial x^i) = 0,
\end{equation}
where
\begin{equation}
\label{eq:H}
H(x^i,p_i) \equiv \frac12 D^{ij}(\bbox {x})p_i p_j + u^i(\bbox {x}) p_i
\end{equation}
is a so-called Wentzell-Freidlin Hamiltonian~\cite{VF}.
Equation~(\ref{eq:eikonal}) has the form of a Hamilton-Jacobi equation,
with $W$~a classical action at zero energy.  To~compute $W(\bbox {x})$ one
may simply use Hamilton's equations of motion to generate the zero-energy
classical trajectory from $S$ to~$\bbox {x}$.  The quantity
$W(\bbox {x})$ will necessarily equal $\int \bbox {p}\cdot d\bbox {x}$, the
line integral being taken from $S$ to~$\bbox {x}$ along the
trajectory. We~stress that $\bbox {p} = \bbox{\nabla} W$ here is not a
physical momentum: it~is a mathematical artifact.  Hamilton's equation
$\dot x^i=D^{ij}p_j + u^i$ reveals that $\bbox {p}$~measures the extent
to which the classical trajectories move against the drift~$\bbox {u}$.
Deterministic (no-noise) trajectories have $\bbox {p}\equiv\bbox{0}$.

By~separating out the ${\cal O}(\epsilon^0)$ terms in ${\cal L}^*_\epsilon
\rho_0=0$ one can show that the pre-exponential factor $K(\bbox {x})$
satisfies~\cite{Naeh90,Talkner87}
\begin{equation}
\label{eq:K}
\dot K=-\left(\bbox{\nabla}\cdot\bbox {u}+D^{ij}W_{,ij}/2\right) K,
\end{equation}
the time derivative being a derivative with respect to transit time along
the zero-energy classical trajectory.  Here $W_{,ij}\equiv \partial
W/\partial x^i \partial x^j$.  By~differentiating the Hamilton-Jacobi
equation one can show that the matrix $(W_{,ij})$ satisfies a Riccati
equation along the trajectory:
\begin{equation}
\label{eq:Riccati}
\dot W_{,ij}=-D^{kl}W_{,ki}W_{,lj}-u^k{}_{,i}W_{,kj}
-u^k{}_{,j}W_{,ki}-u^l{}_{,ij}p_l.
\end{equation}
This facilitates the computation of~$K$.

The zero-energy classical trajectories emanating from the attractor,
sometimes called {\em optimal\/} trajectories, have a direct physical
interpretation: they are the most probable fluctuational trajectories.
If~a noise-induced fluctuation from $S$ to~$\bbox {x}$ occurs, in the
limit of weak noise it~should occur with increasing likelihood along an
optimal trajectory terminating at~$\bbox {x}$.  Such trajectories have
been seen experimentally~\cite{Dykman92}. In~the weak-noise limit the most
probable escape path (MPEP) will be the least-action optimal trajectory
extending from $S$ to the limit cycle~$\partial \Omega$.  Normally this
trajectory will spiral into~$\partial \Omega$, rather than crossing
$\partial\Omega$ in finite time, for the following reason.  If~the MPEP
crossed the limit cycle, the crossing point would be a `hot spot' through
which escape would preferentially occur.  The tangential derivative
$\partial_t W$ ({\em i.e.},~the tangential momentum~$p_t$) would necessarily be
zero there.  But the normal drift velocity~$u^n$ equals zero
on~$\partial\Omega$.  So~the second term in eq.~(\ref{eq:H}) would vanish
at the hotspot.  If~$\bbox{D}$ is nondegenerate, 
eqs.\ (\ref{eq:eikonal})--(\ref{eq:H}) imply that $\bbox{p}=\bbox{0}$ there,
{\em i.e.}, $\dot{\bbox{x}} = \bbox{u}$.  That~is, the ostensible MPEP would be
a deterministic trajectory, which is impossible.  The MPEP normally spirals
into the unstable limit cycle even when $\bbox{D}$ is degenerate.
In~Fig.~\ref{fig:trajs} we~show the MPEP of the van~der Pol model
(\ref{eq:vanderPol1})--(\ref{eq:vanderPol2}).

{\em Multivaluedness\/}.---Dykman, Millonas, and Smelyanskiy
\cite{Dykman94} and the present authors~\cite{MaierCF} have stressed
that $W$ and~$K$ may be multivalued functions of the system state~$\bbox
{x}$, since any given point~$\bbox{x}$ may be the endpoint of more than one
optimal trajectory.  This normally happens in models with an unstable limit
cycle, as Fig.~\ref{fig:trajs} shows.  Optimal trajectories that are
perturbations of the MPEP do not spiral into the limit cycle.  Rather, they
approach~it, and wind around the region~$\Omega$ a~number of times, all the
while deviating farther from the MPEP\null.  They eventually exit
from~$\Omega$ (if~the perturbation is in the outward direction) or move
back toward the attractor (if~the perturbation is inward).  As a
consequence, any point~$\bbox{x}$ near the unstable limit cycle is the
endpoint of any of an {\em infinite\/}, discrete set of optimal
trajectories, which differ from each other in their winding number~$l$,
which may be arbitrarily large.  $W$~and~$K$ are {\em infinite-valued\/},
and the WKB approximation~(\ref{eq:WKB}) generalizes to
\begin{equation}
\label{eq:WKB2}
\rho_0 (\bbox{x}) \sim \sum_l  K^{(l)}(\bbox{x}) \exp\left(- 
W^{(l)}(\bbox{x})/\epsilon\right).
\end{equation}
In~the weak-noise limit, this sum is dominated by the term with minimum
$W^{(l)}(\bbox{x})$.  Equivalently, fluctuations to any point $\bbox
{x}$ in~$\Omega$, in~the limit of weak noise, proceed preferentially along
the {\em physical\/} optimal trajectory from the attractor to~$\bbox{x}$:
the least-action one.  Subdominant trajectories contribute at larger noise
strengths, however.

We shall use (\ref{eq:WKB2}) to compute the flux of Brownian particles over
the barrier~$\partial\Omega$.  We~first approximate $W^{(l)}$ and~$K^{(l)}$
near~$\partial\Omega$ by extending the results of Naeh {\em
et~al.}~\cite{Naeh90}.  As~a first attempt, suppose that $W$~is
single-valued and (to~leading order) quadratic near~$\partial\Omega$,
so~that it~can be approximated as $\Delta W + W_{,nn} n^2/2$.  Here $n$~is
the distance~in from~$\partial\Omega$, in~the normal direction, and the
second normal derivative $W_{,nn}=\partial p_n/\partial n < 0$ depends on
position along~$\partial\Omega$.  The matrix equation~(\ref{eq:Riccati})
yields
\begin{equation}
\label{eq:Riccati2}
\dot W_{,nn} = - D^{nn} \left(W_{,nn}\right)^2 - 2 {u^n}_{,n} W_{,nn}\quad,
\end{equation}
a Riccati equation along~$\partial\Omega$.  The final  term
in~(\ref{eq:Riccati}) has dropped~out, as $\bbox{p} = \bbox{\nabla} W$ is
zero on~$\partial\Omega$ if~$W$ behaves quadratically there.
$W_{,nn}$~as a function of position along~$\partial\Omega$ may
be computed from~(\ref{eq:Riccati2}) by integration~\cite{Naeh90}.

\begin{figure}
\centering
\epsfxsize=2.3in
\leavevmode\epsfbox{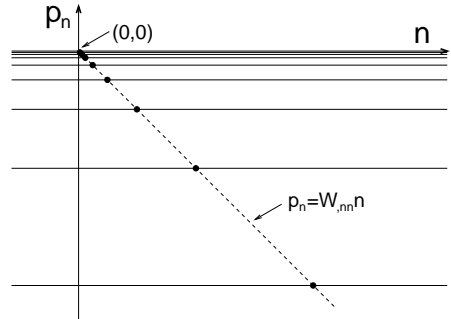}
\caption{A Poincar\'e section.  This sketch shows the points $(n,p_n)$
generated by the optimal trajectories passing~by some specified point
on~$\partial\Omega$.  The dots are generated by the MPEP\null, spiralling
into~$\partial\Omega$.  Cf.\ Figs.\ 1--3 of Graham and
T\'el~\protect\cite{Graham85}.}
\label{fig:section}
\end{figure}

A problem with this approach was pointed out by Graham and
T\'el~\cite{Graham84,Graham85}.  Assuming that $W$~is single-valued near
the unstable limit cycle is much the same as assuming that optimal
trajectories that are perturbations of the MPEP\null, as~well as the MPEP
itself, spiral into the limit cycle.  What actually happens
near~$\partial\Omega$ is revealed by a Poincar\'e section.  Suppose we
choose some point on $\partial\Omega$, and plot the pair $(n,p_n)$, {\em
i.e.}, normal displacement and normal momentum, for each optimal trajectory
that passes nearby.  If~$W$ were quadratic in~$n$, {\em i.e.},
$p_n=\partial W/\partial n$ were linear in~$n$, the points $(n,p_n)$ would
lie on a line with slope~$W_{,nn}$ passing through~$(0,0)$.  What happens
instead is shown in Fig.~\ref{fig:section}.  The MPEP generates points that
tend to~$(0,0)$ geometrically, and lie along the dashed line $p_n=
W_{,nn}n$.  But perturbations of~it generate points that lie along the
horizontal solid lines.

Figure~\ref{fig:section} can be interpreted in~terms of a `return map' that
updates $(n,p_n)$ whenever an optimal trajectory winds once
around~$\Omega$.  This map will have $(0,0)$ as fixed point.  For the MPEP
to spiral into $\partial\Omega$ and yield points along the ideal line, the
linearized return map at~$(0,0)$ must have $(1,W_{,nn})$ as an eigenvector,
with eigenvalue less than~1.  And since deterministic
($\bbox{p}\equiv\bbox{0}$) trajectories that `peel~off'
from~$\partial\Omega$ do~so geometrically, $(1,0)$~must also be an
eigenvector, with eigenvalue greater than~1.  By~Liouville's Theorem
these eigenvalues must be reciprocals, so we denote them $c^{-1}$ and~$c$.
With each turn, the MPEP decreases its distance from~$\partial\Omega$ by a
factor~$c$, and deterministic trajectories that diverge
from~$\partial\Omega$ increase their distance from~it by a factor~$c$.

Figure~\ref{fig:section} can now be explained.  Suppose the MPEP intersects
the $(n,p_n)$ plane at $a(1,W_{,nn})$.  Optimal trajectories that are small
perturbations of the MPEP will intersect~it at $a(1,W_{,nn}) +
\lambda\bbox{v}$, where $\lambda$~is the perturbation strength and
$\bbox{v}$ is model-dependent.  We~write $\bbox{v} = \alpha_s
(1,W_{,nn}) + \alpha_u (1,0)$, where $\alpha_s,\alpha_u\neq0$ in~general.
By~iterating the return map, we~see that after winding $l$ more times, the
trajectories intersect the $(n,p_n)$ plane at
\begin{equation}
\label{eq:horiz}
ac^{-l} (1,W_{,nn}) + \lambda \alpha_u c^l (1,0).
\end{equation}
The $\alpha_s$ term has been dropped here, since it becomes negligible with
respect to the $\alpha_u$ term as~$l\to\infty$.  It~is the second term
in~(\ref{eq:horiz}) that gives rise to the horizontal solid lines of
Fig.~\ref{fig:section}, as $\lambda$ is varied away from zero.

On the MPEP\null, the nonequilibrium potential $W$~behaves quadratically
near~$\partial\Omega$.  In~particular, at~${n=ac^{-l}}$, ${W\approx\Delta W +
W_{,nn}(ac^{-l})^2/2}$.  It~follows that the $l$'th value $W^{(l)}(n)$ of
the infinite-valued function $W(n)$, which arises from trajectories that
wind $l$~times around~$\partial\Omega$, is
\begin{eqnarray}
\label{eq:linearW}
W^{(l)}(n) &\approx & W(ac^{-l}) + (\partial W/\partial n) \left(n - ac^{-l}\right)\\
&=& \Delta W + W_{,nn} (ac^{-l})^2/2 + a c^{-l} W_{,nn} \left(n-ac^{-l}\right)\nonumber\\
&=& \Delta W + W_{,nn} \left(ac^{-l} n - a^2 c^{-2l}/2 \right).
\nonumber
\end{eqnarray}
Each $W^{(l)}$, as a function of the normal distance~$n$, is to leading
order linear, not quadratic.  This has been noticed by Graham and
T\'el~\cite{Graham84,Graham85}, who also noted that if one plots the
physical ({\em i.e.}, minimum) value of~$W(n)$, one obtains a piecewise linear
approximation to the ideal parabola $\Delta W + W_{,nn}n^2/2$.  The
physical~$W$ is nondifferentiable at a sequence of points converging
to~$n=0$.

{\em Oscillatory Asymptotics\/}.---To apply the Kramers method, we~need the
prefactors $K^{(l)}$, as~well as~$W^{(l)}$.  At any~$\bbox{x}$,
$K^{(l)}$~is computed by integrating eq.~(\ref{eq:K}) along an optimal
trajectory that winds $l$~times around~$\Omega$, and terminates
at~$\bbox{x}$.  We~must distinguish here between the `ideal' $W_{,nn}$,
which is a mathematical abstraction (the periodic solution of the Riccati
equation~(\ref{eq:Riccati2})), and the actual second derivatives
$\partial^2 W^{(l)}/\partial x^i\partial x^j$.  It~is the latter that
appear in~(\ref{eq:K}).  In~both (\ref{eq:linearW}) and
Fig.~\ref{fig:section}, which were computed on the basis of the linearized
return map, $\partial^2 W^{(l)}/\partial n^2\equiv 0$, and hence
$\partial^2 W^{(l)}/\partial x^i\partial x^j\equiv 0$, for every~$l$.
Keeping higher-order terms would keep the second derivatives $\partial^2
W^{(l)}/\partial x^i\partial x^j$ from being identically zero, but they
would still fall to zero as $\partial\Omega$~is approached.

It~follows that when computing $K$ near~$\partial \Omega$, we~may
replace~(\ref{eq:K}) by $\dot K\approx
-\left(\bbox{\nabla}\cdot\bbox{u}\right) K$.  In~the limit of large winding
number~$l$, which involves integration along a trajectory that spirals ever
closer to~$\partial \Omega$, this yields $K^{(l+1)}/K^{(l)} \sim
\exp[-\oint (\bbox{\nabla}\cdot\bbox {u})\,dt]$, the integral being taken
once around~$\partial\Omega$.  By~examination, this limiting quotient
equals~$c^{-1}$.  We~shall write $K^{(l)}\approx Ac^{-l}$, where $A$~is a
function of position along~$\partial\Omega$.  This $n$-independent
approximation is increasingly accurate as~$n\to0$.

Substituting the approximations for $K^{(l)}$ and~$W^{(l)}$
into~(\ref{eq:WKB2}) yields an approximation to the steady-state
probability density near the unstable limit cycle, {\em i.e.},
\begin{displaymath}
\rho_0(n) \!\sim\! Ae^{-\Delta W/\epsilon}\!
\sum_l \! c^{-l} \exp\!\left\{ -W_{,nn}\!\left[ac^{-l} n - a^2 c^{-2l}/2\right]\!/\epsilon\right\}
\end{displaymath}
(Recall that $W_{,nn}<0$.)  It~follows that $\bbox {J}[\rho_0]\cdot
\bbox{n}$, the normal component of the probability flux density at
${n=0}$ ({\em i.e.}, through~$\partial\Omega$), is to leading order
\begin{displaymath}
\mbox{const}\times e^{-\Delta W/\epsilon}
\sum_l c^{-2l} \exp \left(-a^2\left|W_{,nn}\right|c^{-2l}/2\epsilon\right).
\end{displaymath}
The sum over winding number~$l$ may be approximated by a discrete analogue
of Laplace's method.  As~$\epsilon\to0$,
the dominant terms in the sum have $l\approx \left|\log
\epsilon\right|/2\log c$.  Let $l^*$ be the greatest integer less than or
equal to $\left|\log \epsilon\right|/2\log c$, and let $h\equiv \left|\log
\epsilon\right|/2\log c - l^*$.  Changing the summation variable to
$k=l-l^*$ allows one to approximate $\bbox {J}[\rho_0]\cdot \bbox{n}$ in
the $\epsilon\to0$ limit, up~to a constant factor, by
\begin{displaymath}
\epsilon^q e^{-\Delta W/\epsilon}
\sum_{k=-\infty}^\infty c^{-2(k-h)} 
\exp \left(-a^2\left|W_{,nn}\right|c^{-2(k-h)}/2\right).
\end{displaymath}
Here $q\equiv{\left(1+2\log c\right)/2}$, and the summation is
periodic in~$h$ with period unity.  Equivalently, the summation is periodic
in~$\left|\log\epsilon\right|$ with period $2\log c$.  Substituting this
flux density into~(\ref{eq:flux}), which involves an integral
over~$\partial\Omega$, yields $R\sim\mbox{const}\times\epsilon^b e^{-\Delta
W/\epsilon} G\left(\left|\log\epsilon\right|\right)$, where the exponent
$b\equiv q-1$, and the function $G(\bullet)$ must have period $2\log c$.
This is the promised oscillatory rate formula.  Interestingly, $b$~varies
continuously as the model is changed.

{\em Discussion\/}.---The slowly oscillating factor
$G\left(\left|\log\epsilon\right|\right)$ is related to a phenomenon
discussed elsewhere~\cite{MaierB}.  If~the noise strength~$\epsilon$ is
small, escape across a quadratic barrier follows the formally most probable
escape path (the MPEP) only until it~gets within an ${\cal
O}(\epsilon^{1/2})$ distance of the barrier.  Thereafter escape occurs
diffusively, rather than ballistically.  Since the MPEP in the models
considered here spirals geometrically into the barrier~$\partial\Omega$,
the point at~which it gets within an ${\cal O}(\epsilon^{1/2})$ distance
will cycle around~$\Omega$ as~$\epsilon\to0$, periodically
in~$\left|\log\epsilon\right|$.  (Cf.~Day~\cite{Day93,Day93a}.)  In~fact,
the period will be $2\log c$.  If~the effective diffusivity varies with
position along~$\partial\Omega$, one would expect the escape rate~$R$ to be
periodically modulated.  That is what we have shown to occur.

We expect the phenomenon of slow oscillations is relevant to stochastic
resonance in multistable continuous systems.  Hu~Gang {\em
et~al.}~\cite{Gang96} have recently considered such systems, with the
addition of time-periodic forcing and external noise.  Steady states are
then periodic attractors, separated by unstable limit cycles.  In~the
weak-noise limit, the rate of noise-induced transitions should therefore
include an oscillatory factor.

This research was partially supported by the National Science Foundation
under grants NCR-90-16211 and DMS-95-00792 (RSM), and by the
U.S. Department of Energy under contract DE-FG03-93ER25155 (DLS).





\end{document}